\documentclass{article}%
\usepackage{amsfonts}
\usepackage{amsmath}
\usepackage{amssymb}
\usepackage{graphicx}
\usepackage[numbers,sort&compress]{natbib}
\usepackage{color}%
\providecommand{\U}[1]{\protect\rule{.1in}{.1in}}
\providecommand{\U}[1]{\protect\rule{.1in}{.1in}}
\newtheorem{theorem}{Theorem}

\newtheorem{example}[theorem]{Example}

\newtheorem{remark}[theorem]{Remark}

\newenvironment{proof}[1][Proof]{\noindent\textbf{#1.} }{\ \rule{0.5em}{0.5em}}
\DeclareMathOperator{\diag}{diag}
\begin{document}

\title{Separable quantizations of St\"{a}ckel systems}
\author{Maciej B\l aszak\\Faculty of Physics, Division of Mathematical Physics, A. Mickiewicz University\\Umultowska 85, 61-614 Pozna\'{n}, Poland\\blaszakm@amu.edu.pl
\and Krzysztof Marciniak\\Department of Science and Technology \\Campus Norrk\"{o}ping, Link\"{o}ping University\\601-74 Norrk\"{o}ping, Sweden\\krzma@itn.liu.se
\and Ziemowit Doma\'{n}ski\\Center for Theoretical Physics of the Polish Academy of Sciences\\Al. Lotnik\'{o}w 32/46, 02-668 Warsaw, Poland\\domanski@cft.edu.pl}
\maketitle

\begin{abstract}
In this article we prove that many Hamiltonian systems that can not be
separably quantized in the classical approach of Robertson and Eisenhardt can
be separably quantized if we extend the class of admissible quantizations
through a suitable choice of Riemann space adapted to the Poisson geometry of
the system. Actually, in this article we prove that for every quadratic in
momenta St\"{a}ckel system (defined on $2n$ dimensional Poisson manifold) for
which St\"{a}ckel matrix consists of monomials in position coordinates there
exist infinitely many quantizations - parametrized by $n$ arbitrary functions
- that turn this system into a quantum separable St\"{a}ckel system.

\end{abstract}


\section{Introduction}

In classical mechanics the Hamiltonian equations of motion are represented by
a system of nonlinear ODE's and are in general not integrable. A famous
exception is the class of the so called Liouville integrable systems, i.e.
those Hamiltonian systems which possess a sufficient number of global
constants of motion in involution. In order to integrate such a system by
quadratures it is necessary to find a distinguish orthogonal coordinates, so
called separation coordinates. Once we find separation coordinates we can
linearize equations of motion according to Hamilton-Jacobi method and then
integrate them. Particular important class of separable systems, specially
from the physical point of view, is represented in literature by so called
St\"{a}ckel systems, with Hamiltonian and all constants of motion quadratic in
momenta. In the present paper we also restrict ourselves to such class of systems.

This paper deals with admissible quantizations of classical St\"{a}ckel
systems and investigation of their quantum integrability and quantum
separability. Surprisingly, in spite of the fact that there exists an
extensive literature on that subject, nevertheless the foundations of the
theory have been formulated in the early 1930's by Robertson and Eisenhart
(see the next section) and have not been changed until now. In their approach
is considered only one particular way of quantization, which we now call
\emph{natural minimal quantization}, i.e. the minimal quantization generated
by the metric from the kinetic part of the Hamiltonian of the system. One of
the results of this classical theory is the so called \emph{Robertson
condition}, the fullfilment of which guarantees the quantum separability of
the stationary Schr\"{o}dinger equation generated by the corresponding
quantized Hamiltonian. In consequence, according to Robertson-Eisenhart
theory, there is only a very limited class of St\"{a}ckel systems which are
quantum separable.

In this paper we broaden the theory by considering quantizations related to
arbitrary metric tensor, not necessarily related with the Hamiltonian of the
system. As a consequence of this new approach, we are able to formulate the
following conjecture:
\begin{align*}
& \text{\emph{For arbitrary St\"{a}ckel system with all constants of motion
quadratic in momenta }}\\
& \text{\emph{there exists a family of quantizations preserving quantum
separability.}}%
\end{align*}

In this paper we prove that conjecture for a very large class of St\"{a}ckel
systems, generated by separation relations of the form (\ref{St}), where
St\"{a}ckel matrix consists of monomials in position coordinates. For any
St\"{a}ckel system from this class we construct a family of metrices for which
the minimal quantization leads to quantum separability and commutativity of
the quantized constants of motion. We want to stress, however, that we do not
deal with spectral theory of the obtained quantum systems, as it requires a
separate investigations.

The paper is organized as follows. In Section \ref{Pre} we briefly summarize
the results of Robertson-Eisenhart theory of quantum separability. In Section
\ref{CS} we present some fundamental facts about classical St\"{a}ckel
systems. Section \ref{aq} contains presentation of some results derived from
our general theory of quantization of Hamiltonian systems on phase space;
especially we demonstrate how to obtain the minimal quantization (\ref{13})
from our general theory. In Section \ref{MinQ} we relate quantizations of the
same Hamiltonian in different metrics $g$ and $\bar{g}$ (or in different
Hilbert spaces $L^{2}(Q,\omega_{g})$ and $L^{2}(Q,\omega_{\bar{g}})$).
Essentially, this construction explains the origin of the quantum correction
terms in the classical Hamiltonians introduced in \cite{Hiet1} and in
\cite{Hiet2}. Section \ref{SQS} is devoted to the issue of separable
quantizations of St\"{a}ckel systems. We construct a family of metric tensors
which fulfill the so called generalized Robertson condition introduced in our
previous paper \cite{bletal2}. Using this condition we prove (Theorem
\ref{wazny}) that there exists an infinite family, parametrized by $n$
arbitrary functions of one variable, of separable quantizations of a given
St\"{a}ckel system from our considered class. Finally, in Section \ref{QIS} we
address the issue of quantum integrability of St\"{a}ckel systems. This
section generalizes in an essential way the results from \cite{Ben2}. We
present the construction of commuting self-adjoint operators in arbitrary
Hilbert spaces $L^{2}(Q,\omega_{g})$, once we have a quantum separable
St\"{a}ckel system. It also contains two illustrative examples. An invariant
form of Theorem \ref{main1} is proved in Appendix.

\section{Preliminaries - legacy of Robertson and Eisenhardt \label{Pre}}

This paper addresses the issue of separable and integrable quantizations of
commuting sets of quadratic in momenta Hamiltonians of the form%
\begin{equation}
H(x,p)=\frac{1}{2}A^{ij}(x)p_{i}p_{j}+V(x)\label{10}%
\end{equation}
(throughout the whole article we apply - unless explicitly stated otherwise -
the Einstein summation convention) defined on a cotangent bundle to some
$n$-dimensional Riemannian manifold $Q$ equipped with metric tensor $g$. The
variables $x=(x_{1},\ldots,x_{n})$ are coordinates on $Q$ and $p_{i}$
conjugate momenta (fiber coordinates in $T^{\ast}Q$) while $A^{ij}(x)$ are
components of a symmetric $(2,0)$-tensor $A$ on $Q$. Note that we do not
assume here any relation between the tensor $A $ and the metric tensor $g$.
The real function $V(x)$ is called the potential of the Hamiltonian
(\ref{10}). Two important partial differential equations can be associated
with the Hamiltonian (\ref{10}): the Hamilton-Jacobi equation%
\begin{equation}
H\left(  x_{1},\dotsc,x_{n},\frac{\partial W}{\partial x_{1}},\dotsc
,\frac{\partial W}{x_{n}}\right)  =a\label{11}%
\end{equation}
for the generating function $W(x,a)$ for a canonical transformation
linearizing the flow of Hamilton equations%
\[
x_{i,t}=\frac{\partial H}{\partial p_{i}},\quad p_{i,t}=-\frac{\partial
H}{\partial x_{i}},\quad i=1,\dotsc,n
\]
(here and in what follows the comma denotes the differentiation with respect
to a variable) associated with (\ref{10}), and the stationary Schr\"{o}dinger
equation%
\begin{equation}
\hat{H}\Psi(x)=E\Psi(x)\label{12}%
\end{equation}
where%
\begin{equation}
\hat{H}=-\frac{\hbar^{2}}{2}\nabla_{i}A^{ij}\nabla_{j}+V(x)\label{13}%
\end{equation}
is the Hamilton operator (quantum Hamiltonian) acting on the Hilbert space
$L^{2}\left(  Q,\left\vert \det g\right\vert ^{1/2}dx\right)  $ of square
integrable (in the measure $\omega_{g}=\left\vert \det g\right\vert ^{1/2}dx
$) complex functions on $Q$. The operators $\nabla_{i}$ are operators of
Levi-Civita connection associated with the metric $g$ and $\hbar$ is the
Planck constant. One says then that the Hamilton operator (\ref{13}) is the
\emph{quantization} of the Hamiltonian (\ref{10}) in the metric $g$. Note that
the above quantization procedure is so far defined ad hoc, arbitrarily.

An important issue related with equations (\ref{11})\ and (\ref{12}) is the
problem of their \emph{separability}. We say that the Hamilton-Jacobi equation
(\ref{11}) is additively separable if it admits a solution
\begin{equation}
W(x,a)=\sum\limits_{i=1}^{n}W_{i}(x_{i},a)\label{14}%
\end{equation}
depending in a suitable manner on $n$ additional parameters $a=(a_{1}%
,\ldots,a_{n})$ (the solution (\ref{14}) is often called a \emph{complete
integral} of (\ref{11})). Similarly, we say that\ the Schr\"{o}dinger equation
(\ref{12}) is multiplicatively separable if it admits a solution
\begin{equation}
\Psi(x,a)=\prod_{i=1}^{n}\psi_{i}(x_{i},a)\label{15}%
\end{equation}
depending in a suitable way on $2n$ additional parameters $a=(a_{1}%
,\ldots,a_{2n})$. P. St\"{a}ckel showed in \cite{Stackel} the necessary and
sufficient conditions for separability of (\ref{10}) in orthogonal (with
respect to $A$) coordinates (meaning that $A$ has to be diagonal in the
variables $x$). Assume thus that $A$ plays the role of the contravariant
metric (i.e. that $A=G$, where $G=g^{-1}$) and that the metric $G$ is diagonal
in coordinates $x$. Robertson \cite{rob27} proved that in this case if the
Hamilton-Jacobi equation (\ref{11}) separates in the variables $x$ then the
Schr\"{o}dinger equation (\ref{12}) also separates provided that an additional
condition, called today Robertson condition, is satisfied. Eisenhart in
\cite{eis34} proved that Robertson condition is satisfied if and only if the
Ricci tensor $R_{ij}$ of the metric $g\,\ $is diagonal. We stress again that
in these works $A=G$ and in this particular case the Hamilton operator
(\ref{13}) takes the form
\[
\hat{H}=-\frac{\hbar^{2}}{2}G^{ij}\nabla_{i}\nabla_{j}+V(x)
\]
Robertson actually claimed in his theorem that the separability of
Schr\"{o}dinger equation also implies separability of Hamilton-Jacobi
equation; this statement is not correct if we use the definition of
separability used by Robertson. Benenti et al in \cite{Ben1} completed the
works of Robertson and Eisenhart by introducing an appropriate definition of
separability of Schr\"{o}dinger equation, involving $2n$ parameters $a_{i}%
\,$\ as in (\ref{15}) (Robertson had no parameters in his definition of
separability, a drawback not observed by Eisenhart). Assuming the definition
of Benenti et al the theorem of Robertson becomes:

\begin{theorem}
Assume that $A=G$ and that $G$ is diagonal in the variables $x_{i}$. The
Schr\"{o}dinger equation (\ref{12}) admits a separable solution (\ref{15}) if
and only if the Hamilton-Jacobi equation (\ref{11}) admits a separable
solution (\ref{14}) and moreover if the Robertson condition
\begin{equation}
R_{ij}=0\text{ for all }i\neq j\label{rob1}%
\end{equation}
is satisfied.
\end{theorem}

One can show that in orthogonal coordinates%
\begin{equation}
R_{ij}=\frac{3}{2}\partial_{i}\Gamma_{j},\ \ \ i\neq j\label{rij}%
\end{equation}
where $\Gamma_{i}$ are metrically contracted Christoffel symbols of $g$
defined by
\begin{equation}
\Gamma_{i}=g_{il}G^{jk}\Gamma_{jk}^{l}\text{, \ \ }i=1,\ldots,n\label{mcCh}%
\end{equation}
Thus, in orthogonal coordinates the Robertson condition becomes%
\begin{equation}
\partial_{i}\Gamma_{j}=0\text{ for }j\neq i\label{rob}%
\end{equation}

In papers \cite{rob27} and \cite{eis34} the authors considered a quantization
procedure for only one Hamiltonian and assumed that the underlying metric of
the configuration space is defined by the tensor $A$ in the Hamiltonian, i.e.
they assumed that $A=G$. Suppose now that we have $n$ ($n=\dim Q$)
Poisson-commuting (so they constitute an integrable system in the sense of
Liouville) Hamiltonians each of the form (\ref{10}):%
\begin{equation}
H_{r}=\frac{1}{2}A_{r}^{ij}p_{i}p_{j}+V_{r}(x),\quad r=1,\dotsc,n.\label{16}%
\end{equation}
A natural question one can pose is whether the corresponding quantum
Hamiltonians $\hat{H}_{r}$ (acting in the Hilbert space $L^{2}\left(
Q,\omega_{g}\right)  $ defined by the metric $G=A_{1}$) will constitute a
quantum integrable systems i.e. whether they will commute. In \cite{Ben2} the
authors proved that this happens if and only if the so called pre-Robertson
condition%
\begin{equation}
\partial_{i}R_{ij}-\Gamma_{i}R_{ij}=0,\quad i\neq j\label{prerob1}%
\end{equation}
is satisfied. Due to (\ref{rij}), this condition in orthogonal coordinates
reads%
\begin{equation}
\partial_{i}^{2}\Gamma_{j}-\Gamma_{i}\partial_{i}\Gamma_{j}=0,\quad i\neq
j.\label{prerob}%
\end{equation}

\begin{remark}
The Robertson condition (\ref{rob1}) or (\ref{rob}) implies the pre-Robertson
condition (\ref{prerob1}) or (\ref{prerob}) so quantum separability implies
the quantum integrability, as it is in the classical case.
\end{remark}

The above theory describes the quantization of a Hamiltonian, or a set of
Hamiltonians, of the form (\ref{10}) in the case when one of the tensors
$A_{r}$ plays the role of the metric. However, Hamiltonians are functions on a
phase space with no obvious metric given. In this paper we will therefore
develop the theory of quantization of Hamiltonians of type (\ref{10}) in
Hilbert spaces $L^{2}(Q,\omega_{g})$ defined by the metric not related to
these Hamiltonians. Let us thus pose the following question: given a separable
Hamiltonian system consisting of $n$ Hamiltonians of the form (\ref{16}), how
to find metric tensor(s) in which an appropriate quantization procedure turns
this system into a separable and integrable quantum system? We will answer
this question in the spirit of papers
\cite{Blaszak:2012,Blaszak:2013b,Blaszak:2014} where we have developed a
general theory of quantizing Hamiltonian systems directly on the phase space;
the quantization in this approach is given by an appropriate deformation of
Poisson algebra of classical observables (real functions) on the phase space
$M$ to a quantum algebra. Various deformations of this algebra are related to
each other by an automorphism $S$. However, in order to make this article as
compact as possible, we will almost completely omit this general setting but
use its results in the position representation, that is, we will work directly
in Hilbert spaces $L^{2}(Q,\omega_{g})$ of the functions defined on the base
manifold $Q.$

\section{Classical St\"{a}ckel systems in separation coordinates and adapted
Riemannian geometry\label{CS}}

Consider a $2n$-\hspace{0pt}dimensional connected Poisson manifold
$(M,\mathcal{P})$, where $\mathcal{P}$ is a non-degenerated Poisson tensor. An
integrable system is a set of $n$ real valued functions $H_{i}$ on $M$ in
involution with respect to a Poisson bracket:
\[
\{H_{i},H_{j}\}:=\mathcal{P}(dH_{i},dH_{j})=0,\quad i,j=1,\dotsc,n.
\]
The functions $H_{i}$ generate $n$ pairwise commuting Hamiltonian equations
\begin{equation}
u_{,t_{i}}=\mathcal{P}d{H_{i}},\quad i=1,\dotsc,n,\quad u\in M.\label{int}%
\end{equation}
i.e. an integrable system. Let us fix a set $(x,p)=(x_{1},\ldots x_{n}%
,p_{1},\ldots p_{n})$ of Darboux (canonical) coordinates on $M$ (so that
$\{x_{i},x_{j}\}=\{p_{i},p_{j}\}=0$, $\{x_{i},p_{j}\}=\delta_{ij}$). One of
the methods of solving the equations (\ref{int}) is to find a solution
$W(x,a)$ to the system of Hamilton-Jacobi equations (\ref{11}) corresponding
to the Hamiltonians $H_{i}$
\begin{equation}
H_{i}\left(  x_{1},\dotsc,x_{n},\frac{\partial W}{\partial x_{1}},\dotsc
,\frac{\partial W}{x_{n}}\right)  =a_{i},\quad i=1,\dotsc,n.\label{HJ}%
\end{equation}
The solution $W(x,a)$ is then a generating function for a canonical
transformation $(x,p)\mapsto(b,a)$ to a new set of coordinates on $M$ (with
$a_{i}=H_{i}$) in which the equations (\ref{int}) attain the form%
\[
b_{i,t_{j}}=\delta_{ij},\quad a_{i,t_{j}}=0
\]
so that all the flows in (\ref{int}) linearize in coordinates $(b,a)$. In most
cases the system of PDE's (\ref{HJ}) is a highly nonlinear system that is very
difficult to solve. However, as we mentioned in introduction, a very appealing
situation occurs if we can find Darboux coordinates $(\lambda,\mu
)=(\lambda_{1},\ldots\lambda_{n},\mu_{1},\ldots\mu_{n})$ in which there exists
a complete integral for all the Hamilton-Jacobi equations (\ref{HJ}) of the form%

\[
W(\lambda,a)=\sum\limits_{i=1}^{n}W_{i}(\lambda_{i},a)
\]
(see (\ref{14})) where each function $W_{i}$ depend only on one canonical
coordinate $\lambda_{i}$ and in a nontrivial way on all parameters
$a=(a_{1},\ldots,a_{n})$. In such a case the systems of PDE's (\ref{HJ}) split
into $n$ uncoupled ODE's for the functions $W_{i}$, which makes it possible to
solve them by quadratures. The coordinates $(\lambda,\mu)$ are then called
separation coordinates of the system (\ref{int}).

The most convenient way to obtain separable systems is to define them directly
in separation coordinates. It is done with the help of the so called
separation relations \cite{Sklyanin}, i.e. $n$ algebraic relations of the form%
\begin{equation}
\varphi_{i}(\lambda_{i},\mu_{i},a_{1},\dotsc,a_{n})=0,\quad i=1,\dotsc
,n\label{SR}%
\end{equation}
each depending on one pair of canonical coordinates and on parameters $a_{i}$.
If there exists an open dense set $\Omega\subset$ $M$ on which the relations
(\ref{SR}) can be solved with respect to the coefficients $a_{i}$ yielding%
\[
a_{i}=H_{i}(\lambda,\mu),\quad i=1,\dotsc,n
\]
then it is easy to show that the functions $H_{i}$ Poisson commute (i.e.
constitute a Liouville integrable system as defined above) and moreover that
the coordinates $(\lambda,\mu)$ are separation coordinates for the
Hamiltonians $H_{i}.$

One of the most important classes of separable systems are the so called
St\"{a}ckel systems, introduced by P. St\"{a}ckel in \cite{Stackel} and
thoroughly studied in literature (see for example
\cite{bensol,bensol2,bensol3}). They are generated by separation relations
linear in Hamiltonians $H_{i}$ and quadratic in canonical momenta $\mu_{i}$.
In our paper we restrict ourselves to a --- still very general --- class of
St\"{a}ckel systems defined by the following separation relations
\begin{equation}
H_{1}\lambda_{i}^{\gamma_{1}}+H_{2}\lambda_{i}^{\gamma_{2}}+\dotsb
+H_{n}\lambda_{i}^{\gamma_{n}}=\frac{1}{2}f_{i}(\lambda_{i})\mu_{i}^{2}%
+\sigma_{i}(\lambda_{i}),\quad i=1,\dotsc,n,\label{St}%
\end{equation}
where $\gamma_{i}$ are natural numbers such that $\gamma_{1}>\gamma_{2}%
>\dotsb>\gamma_{n}=0$ (the last choice is for our convenience only) have no
common divisor, and where $f_{i},\sigma_{i}$ are some rational functions of
one argument. The separation relations (\ref{St}) can be written in a matrix
form as
\begin{equation}
S_{\gamma}H=U,\label{vec}%
\end{equation}
where $H=(H_{1},\dotsc,H_{n})^{T}$ and $U=(\frac{1}{2}f_{1}(\lambda_{1}%
)\mu_{1}^{2}+\sigma_{1}(\lambda_{1}),\dotsc,\frac{1}{2}f_{n}(\lambda_{n}%
)\mu_{n}^{2}+\sigma_{n}(\lambda_{n}))^{T}$ is a St\"{a}ckel vector and where
the matrix $S_{\gamma}$ given by%
\[
S_{\gamma}=\left(
\begin{array}
[c]{cccc}%
\lambda_{1}^{\gamma_{1}} & \lambda_{1}^{\gamma_{2}} & \cdots & 1\\
\vdots & \vdots &  & 1\\
\lambda_{n}^{\gamma_{1}} & \lambda_{n}^{\gamma_{2}} & \cdots & 1
\end{array}
\right)
\]
is a particular St\"{a}ckel matrix with functions being monomials parametrized
by the natural numbers $\gamma_{i}$. We can now take as the set $\Omega$ what
remains of $M$ after removing the set of points where $\det S_{\gamma}=0$ as
well as all the poles of $f_{i}$ and $\sigma_{i}$. Solving the relations
(\ref{vec}) on $\Omega$ we obtain the St\"{a}ckel Hamiltonians%

\begin{equation}
H_{r}=\frac{1}{2}\mu^{T}A_{r}\mu+V_{r}(\lambda),\quad r=1,\dotsc,n,\label{Ham}%
\end{equation}
with
\begin{equation}
A_{r}=\diag((S_{\gamma}^{-1})_{r1}f_{1}(\lambda_{1}),\dotsc,(S_{\gamma}%
^{-1})_{rn}f_{n}(\lambda_{n}))\quad\label{Ar}%
\end{equation}
being diagonal matrices with entries that are functions of $\lambda$-variables
only and with the potentials of the form%
\[
V_{r}=\sum_{i}(S_{\gamma}^{-1})_{ri}\sigma_{i}(\lambda_{i})\quad r=1,\dotsc,n.
\]
The systems of the above class, albeit not general St\"{a}ckel systems, still
encompass majority of the St\"{a}ckel systems considered in literature.

Let us now introduce some Riemannian geometry into our considerations. The
specifications below will be motivated by the fact that our quantization
procedure will be performed in appropriate (pseudo\nobreakdash-)Riemannian
spaces. Thus, from now on we will suppose that our manifold $M$ is a cotangent
bundle to some pseudo-Riemannian manifold i.e. $M=T^{\ast}Q$ with $Q$ equipped
with some metric tensor $g$. We will also make three additional assumptions:

\begin{enumerate}
\item The manifold $(Q,g)$ and the Poisson structure are adapted to each other
in the sense that the first $n$ Darboux coordinates $\lambda_{i}$ are
coordinates on $Q$ while the remaining Darboux coordinates $\mu_{i}$ are fiber coordinates.

\item Coordinates $\lambda_{i}$ are orthogonal coordinates for the metric $g $
i.e. $g$ and $G = g^{-1}$ are diagonal (but not necessarily flat) in
$\lambda_{i}$.

\item The base manifold $Q$ is almost covered by a single, open and dense in
$M$, chart with coordinates $(\lambda_{1},\ldots,\lambda_{n})$.
\end{enumerate}

The matrices $A_{r}$ in (\ref{Ham}) can now be interpreted as $(2,0)$-tensors
on $Q$ that can be written as%
\[
A_{r}=T_{r}G,\quad r=1,\ldots,n
\]
where $T_{r}$ are $(1,1)$-tensors on $Q$. Further, in a very special case when
$G=A_{1}$ the tensors $T_{r}$ are Killing tensors for the metric $G$. We will
denote them as $K_{r}$, so that%
\[
A_{r}=K_{r}A_{1},\quad r=1,\ldots,n
\]

A particular subclass of St\"{a}ckel systems (\ref{St}) is then given by
choosing $\gamma_{i}=n-i$. Such systems are called St\"{a}ckel system of
\emph{Benenti type} (or simply \emph{Benenti systems}) and are thus generated
by the separation relations of the form%

\begin{equation}
H_{1}\lambda_{i}^{n-1}+H_{2}\lambda_{i}^{n-2}+\dotsb+H_{n}=\frac{1}{2}%
f_{i}(\lambda_{i})\mu_{i}^{2}+\sigma_{i}(\lambda_{i}),\quad i=1,\dotsc
,n\label{Stb}%
\end{equation}
It can be shown that in the Benenti case the metric tensor $G=A_{1}$ has the
form%
\begin{equation}
A_{1}=\diag\left(  \frac{f_{1}(\lambda_{1})}{\Delta_{1}},\ldots,\frac
{f_{n}(\lambda_{n})}{\Delta_{n}}\right)  ,\quad\Delta_{i}=\prod\limits_{j\neq
i}(\lambda_{i}-\lambda_{j})\label{A1}%
\end{equation}
while the Killing tensors $K_{r}$ are of the form%
\begin{equation}
K_{r}=-\diag\left(  \frac{\partial\rho_{r}}{\partial\lambda_{1}},\cdots
,\frac{\partial\rho_{r}}{\partial\lambda_{n}}\right)  ,\quad r=1,\ldots
,n\label{Kr}%
\end{equation}
with $\rho_{i}=\rho_{i}(\lambda)$ being signed symmetric polynomials
(Vi\`{e}te polynomials) in the variables $\lambda_{1},\ldots,\lambda_{n}$:%
\begin{equation}
\rho_{i}(\lambda)=(-1)^{i}\sum\limits_{1\leq s_{1}<s_{2}<\ldots<s_{i}\leq
n}\lambda_{s_{1}}\ldots\lambda_{s_{i}},\quad i=1,\ldots,n\label{defq}%
\end{equation}

Let us now go back to an arbitrary St\"{a}ckel system of the form (\ref{St})
defined by the choice of the constants $\gamma_{1}>\gamma_{2}>\dotsb
>\gamma_{n}=0$ and the choice of functions $f_{i},\sigma_{i}$. Then the
tensors $A_{r}$ for this system can be written as \cite{bensol}%
\begin{equation}
A_{r}=\frac{1}{\varphi}\chi_{r}G_{B,f},\quad r=1,\ldots,n\label{Are}%
\end{equation}
where $G_{B,f}$ is the corresponding Benenti metric given by (\ref{A1})%
\begin{equation}
G_{B,f}=\diag\left(  \frac{f_{1}(\lambda_{1})}{\Delta_{1}},\ldots,\frac
{f_{n}(\lambda_{n})}{\Delta_{n}}\right) \label{GB}%
\end{equation}
where $\chi_{r}$ are some polynomial functions of the Killing tensors $K_{r} $
in (\ref{Kr}) and where
\begin{equation}
\varphi=\det\left(
\begin{array}
[c]{ccc}%
\rho_{n_{1}-1} & \cdots & \rho_{n_{1}-k}\\
\vdots & \ddots & \vdots\\
\rho_{n_{k}-1} & \cdots & \rho_{n_{k}-k}%
\end{array}
\right) \label{phi}%
\end{equation}
(where we adapt the notation $\rho_{0}=1$ and $\rho_{i}=0$ for $i<0$ or $i>n$)
while the constants $n_{i}$ are those for which the corresponding monomials
$\lambda^{n+k-n_{i}}$ are missing in the left hand side of (\ref{St}) (they
are \textquotedblleft holes\textquotedblright\ in the sequence $\left\{
\gamma_{1}=n+k-1,\gamma_{2},\dotsb,\gamma_{n}=0\right\}  $ numbered from the
left;\thinspace\ \thinspace$k$ is determined from the equation $\gamma
_{1}=n+k-1$). Note that if such \textquotedblleft holes\textquotedblright\ are
absent (as in Benenti case) then $\varphi=1$. For example, if the left hand
side of the St\"{a}ckel system is $H_{1}\lambda^{4}+H_{2}\lambda+H_{3}$, then
$n=3$, $k=2$, $n_{1}=2$, $n_{2}=3$ and the function (\ref{phi}) becomes:%
\[
\varphi=\det\left(
\begin{array}
[c]{cc}%
\rho_{1} & \rho_{0}\\
\rho_{2} & \rho_{1}%
\end{array}
\right)  =\lambda_{1}^{2}+\lambda_{1}\lambda_{2}+\lambda_{1}\lambda
_{3}+\lambda_{2}^{2}+\lambda_{2}\lambda_{3}+\lambda_{3}^{2}%
\]
We finish this chapter with an important remark.

\begin{remark}
\label{uwaga2}If $f_{i}=f$ and if $f$ is a polynomial of order $\leq n$ then
the metric $G_{B,f}$ in (\ref{GB}) is flat.
\end{remark}

\section{Admissible quantizations of quadratic in momenta Hamiltonians on
pseudo-Riemannian spaces\label{aq}}

A usual way of quantization of a given Hamiltonian system living on a phase
space $M=\mathbf{R}^{2n}$ is by replacing the observables of the system (i.e.
real functions on the phase space of the system, written as functions of
positions $x_{i}$ and momenta $p_{i}$) by self-adjoint operators acting on the
Hilbert space $\mathcal{H}=L^{2}(\mathbf{R}^{n})$ of square integrable complex
functions on $\mathbf{R}^{n}$. This is done by replacing $x_{i}$ and $p_{i}$
in the observables by the non-commuting operators $\hat{x}_{j}=x_{j}$ and
$\hat{p}_{j}=-i\hbar\partial/\partial x_{j}$ acting on $L^{2}(\mathbf{R}^{n}%
)$. In this procedure we have to agree on a certain order of non-commuting
operators $\hat{x}_{j}$ and $\hat{p}_{j}$ in the obtained operator. One
usually applies the Weyl ordering that guarantees that the obtained operators
will be self-adjoint.

Suppose now that we want to quantize in a coordinate-free way a Hamiltonian
system given on a phase space $M=T^{\ast}Q$ that is the cotangent bundle to a
pseudo-Riemannian manifold equipped with a metric tensor $g$. In a series of
papers \cite{Blaszak:2012, Blaszak:2013b, Blaszak:2014} we have developed a
consistent theory of quantizing a Hamiltonian system directly on the phase
space $M$ through a very general procedure of quantization. Here we briefly
sketch some parts of this construction that are important for our further
considerations; we perform the construction in the so called position representation.

Let us thus choose a canonical (Darboux) coordinate system $(x_{j},p_{j})$ on
$M$ satisfying assumptions 1--3 from the previous section. Thus, $x_{j}$ are
some coordinates on $Q$ and $p_{j}$ are the corresponding conjugate momenta.
Let us also (following \cite{Blaszak:2013b}, \cite{deWitt}, and \cite{essen})
introduce the operators%
\begin{equation}
\hat{x}_{j}=x_{j},\quad\hat{p}_{j}=-i\hbar\left(  \frac{\partial}{\partial
x_{j}}+\frac{1}{2}\Gamma_{jk}^{k}\right) \label{op}%
\end{equation}
acting on the Hilbert space $\mathcal{H}=L^{2}(Q,\omega_{g})$ of functions on
the base manifold $Q$ (configuration space) where $\omega_{g}=\left\vert \det
g\right\vert ^{1/2}dx$ is a volume form defined by the metric $g$ and where
$\Gamma_{jk}^{k}$ are contracted Christoffel symbols of the metric $g$. The
operators (\ref{op}) are self-adjoint in $\mathcal{H}$ and moreover are
canonical quantum operators as $\left[  \hat{x}_{j},\hat{p}_{k}\right]
=i\hbar\delta_{jk}$.

Now, a given observable $H=H(x,p)$ can be quantized in many different ways by
applying different orderings to the operators $\hat{x},\hat{p}$ in $H(\hat
{x},\hat{p})$. This can be systematically done using a two-parameter family of
automorphisms $S$, introduced in \cite{Blaszak:2014}, acting on the space of
functions on $M$. Any automorphism $S$ from this family relates a given
quantization with a Moyal quantization corresponding to our chosen Darboux
coordinates $(x,p)$.

Our two-parameter family of automorphisms $S$ is up to $\hbar^{2}$-terms given
by
\begin{align}
S &  =1+S_{2}\hbar^{2}+o(\hbar^{4})\nonumber\\
&  =1+\frac{\hbar^{2}}{4!}\left[  3(\Gamma_{lj}^{i}\Gamma_{ik}^{l}%
+aR_{jk})\partial_{p_{j}}\partial_{p_{k}}+3\Gamma_{jk}^{i}\partial_{x_{i}%
}\partial_{p_{j}}\partial_{p_{k}}+(2\Gamma_{nl}^{i}\Gamma_{jk}^{n}%
-\Gamma_{jk,l}^{i})p_{i}\partial_{p_{j}}\partial_{p_{k}}\partial_{p_{l}%
}\right. \label{S}\\
&  \quad-3b\partial_{p_{j}}(\partial_{x_{j}}+\Gamma_{jl}^{i}p_{i}%
\partial_{p_{l}})\partial_{p_{k}}(\partial_{x_{k}}+\Gamma_{kn}^{r}%
p_{r}\partial_{p_{n}})]+o(\hbar^{4}),\nonumber
\end{align}
($a$ and $b$ are real parameters and $\Gamma_{jk,l}^{i}=\partial_{x_{l}}%
\Gamma_{jk}^{i}$) with the inverse given formally by%
\begin{equation}
S^{-1}=1-S_{2}\hbar^{2}+o(\hbar^{4})\label{Sinv}%
\end{equation}

\begin{remark}
\label{uwaga}The terms $o(\hbar^{4})$ in (\ref{S}) are at least of the fourth
order in $\partial_{p_{j}}$ so the formulas (\ref{S})-(\ref{Sinv}) are enough
to calculate the action of $S$ respectively $S^{-1}$ on Hamiltonians that are
up to third order in momenta.
\end{remark}

We can now introduce the following quantization procedure of a given
observable $H(x,p)$:

\begin{enumerate}
\item Deformation of $H(x,p)$ to a new function $H^{\prime}(x,p)=S^{-1}H$ by
an automorphism $S$ from our family (\ref{S})

\item Replacing $x_{j}$ and $p_{j}$ in $H^{\prime}(x,p)$ by the operators
(\ref{op}), which yields the operator $H^{\prime}(\hat{x},\hat{p})$

\item Weyl ordering of the obtained operator.
\end{enumerate}

In short, the $S$-quantization of $H(x,p)$ in the metric $g$ is the operator%
\begin{equation}
\hat{H}=(S^{-1}H)_{W}(\hat{x},\hat{p})\label{q}%
\end{equation}
(where $W$ denotes the Weyl ordering) with operators $\hat{x},\hat{p} $ given
by (\ref{op}) and with a chosen automorphism $S$ from our two-parameter
family. It can be shown that this procedure applied to any classical (real)
observable on $M$ yields a self-adjoint operator on $\mathcal{H}%
=L^{2}(Q,\omega_{g})$.

\begin{remark}
The presented procedure is invariant under the canonical change of coordinates
in the sense that if we start from another canonical set of coordinates
satisfying assumptions 1--3 from the previous section we obtain the quantum
operator that is unitarily equivalent to\ $\hat{H}$.
\end{remark}

Applying the above quantization procedure with the automorphism $S$ as in
(\ref{S}) to a quadratic in momenta Hamiltonian
\begin{equation}
H=\frac{1}{2}p^{T}Ap+V(x)\label{Ham2}%
\end{equation}
yields the two-parameter family of operators (quantum Hamiltonians) on
$\mathcal{H}$ \cite{Blaszak:2014}:
\begin{align}
\hat{H}  & =-\frac{\hbar^{2}}{2}\left(  \nabla_{i}A^{ij}\nabla_{j}+\frac{1}%
{4}(1-b)A_{\text{ \ };ij}^{ij}-\frac{1}{4}(1-a)A^{ij}R_{ij}\right)
+V(x)\label{Hs}\\
& =-\frac{\hbar^{2}}{2}\nabla_{i}A^{ij}\nabla_{j}+\hbar^{2}V_{\text{quant}%
}(x)+V(x)\nonumber
\end{align}
where $\nabla_{i}$ is the operator of the covariant derivative of the
Levi-Civita connection defined by $g$, $R_{ij}$ is the Ricci tensor of $g$ and
where the semicolon $;$ denotes the covariant derivative. The term
$V_{\text{quant}}(x)$ can be considered as a \textquotedblleft quantum
correction\textquotedblright\ to the potential $V$ that comes from the
quantization process. All considered in literature quantizations of quadratic
in momenta Hamiltonians can be obtained by choosing appropriate values of $a$
and $b$ in (\ref{Hs}). In the special case when $A=G$ the formula (\ref{Hs})
reduces to%
\begin{equation}
\hat{H}=-\frac{\hbar^{2}}{2}\left(  G^{ij}\nabla_{i}\nabla_{j}-\frac{1}%
{4}(1-a)R\right)  +V(x)\label{Hss}%
\end{equation}
where $R$ is the Ricci scalar. In the flat case (so that $R_{ij}=0$) and with
\thinspace$b=0$ we obtain the Weyl quantization written in a covariant form,
and (\ref{Hs}) and its specification (\ref{Hss}) attain the form%
\[
\hat{H}=-\frac{\hbar^{2}}{2}\left(  \nabla_{i}A^{ij}\nabla_{j}+\frac{1}%
{4}A_{\text{ \ };ij}^{ij}\right)  +V(x)
\]
and%
\[
\hat{H}=-\frac{\hbar^{2}}{2}G^{ij}\nabla_{i}\nabla_{j}+V(x)
\]
respectively.

As we see, in the general quantization scheme there appear the quantum
correction term $V_{\text{quant}}(x)$ to the potential $V$. This quantum
potential is in general non-separable \cite{bletal}, so from the point of
quantum separability the optimal choice of quantization is given \ by $a=b=1$,
which yields%

\begin{equation}
\hat{H}=-\frac{\hbar^{2}}{2}\nabla_{i}A^{ij}\nabla_{j}+V(x)\label{mq}%
\end{equation}
This quantizations is called a \emph{minimal} quantization induced by the
metric tensor \thinspace$g$ and $V_{\text{quant}}(x)=0$ in that case. It is
exactly the a priori quantization considered by Eisenhardt, Robertson, Benenti
and many others and described in the preliminary part above. Our theory
clearly explains its origin and shows that this is but one of infinitely many
possibilities of quantizing the Hamiltonian (\ref{10}).

\section{Minimal quantization in different metric spaces\label{MinQ}}

Our goal now is to relate two minimal quantizations induced by different
metric tensors. We will need this in order to be able to write systems of
commuting operators in various Hilbert spaces with measures induced by
different metrics.

Consider thus two different metric tensors $g$ and $\bar{g}$. As usual, we
will denote their contravariant forms by $G$ and $\bar{G}$, respectively. Each
of these metrics induces a minimal quantization (described in Section~\ref{aq}%
) by morphisms $S$ and $\bar{S}$, respectively, where (cf. (\ref{S}) with
$a=b=1$)%
\begin{align}
S &  =1+\frac{\hbar^{2}}{4!}\left[  3(\Gamma_{lj}^{i}\Gamma_{ik}^{l}%
+R_{jk})\partial_{p_{j}}\partial_{p_{k}}+3\Gamma_{jk}^{i}\partial_{x_{i}%
}\partial_{p_{j}}\partial_{p_{k}}+(2\Gamma_{nl}^{i}\Gamma_{jk}^{n}%
-\Gamma_{jk,l}^{i})p_{i}\partial_{p_{j}}\partial_{p_{k}}\partial_{p_{l}%
}\right. \label{Sg}\\
&  \quad\left.  -3\partial_{p_{j}}(\partial_{x_{j}}+\Gamma_{jl}^{i}%
p_{i}\partial_{p_{l}})\partial_{p_{k}}(\partial_{x_{k}}+\Gamma_{kn}^{r}%
p_{r}\partial_{p_{n}})+o(\hbar^{4})\right]  ,\nonumber
\end{align}
and where $\bar{S}$ is given by an analogous expression with $\Gamma_{jk}^{i}$
replaced by Christoffel symbols $\bar{\Gamma}_{jk}^{i}$ of the Levi-Civita
connection induced by $\bar{g}$. For a (classical) observable of the form
\begin{equation}
H(x,p)=\frac{1}{2}A^{ij}(x)p_{i}p_{j}+V(x)\label{Ht}%
\end{equation}
by (\ref{q}), its minimal quantization with respect to $g$ is given by%
\begin{equation}
\hat{H}=(S^{-1}H)_{W}(\hat{x},\hat{p})=-\frac{\hbar^{2}}{2}\nabla_{i}%
A^{ij}\nabla_{j}+V(x)\label{op1}%
\end{equation}
and acts in $L^{2}(Q,\omega_{g})$, while its quantization with respect to
$\bar{g}$ is given by a similar expression involving $\bar{\nabla}_{i}\,
$\ (that is the covariant differentiation with respect to $\bar{g} $) and the
operators $\hat{\bar{x}}_{j}=x_{j}$ and $\hat{\bar{p}}_{j}=-i\hbar\left(
\partial_{j}+\frac{1}{2}\bar{\Gamma}_{jk}^{k}\right)  $). These are in general
two different operators, acting in two different Hilbert spaces:
$L^{2}(Q,\omega_{g})$ and $L^{2}(Q,\omega_{\bar{g}})$, respectively. The
Hilbert spaces $L^{2}(Q,\omega_{g})$ and $L^{2}(Q,\omega_{\bar{g}})$ are
however isometric, with the isometry $L^{2}(Q,\omega_{g})\rightarrow
L^{2}(Q,\omega_{\bar{g}})$ given by%
\begin{equation}
\bar{\Psi}=U\Psi=\frac{\left\vert \det g\right\vert ^{1/4}}{\left\vert
\det\bar{g}\right\vert ^{1/4}}\Psi\label{I}%
\end{equation}
where $\Psi\in L^{2}(Q,\omega_{g})$ and $\bar{\Psi}\in L^{2}(Q,\omega_{\bar
{g}})$. The isometry (\ref{I}) induces a similarity map\ between operators in
both spaces: it maps an operator $\hat{F}$ acting in $L^{2}(Q,\omega_{g})$ to
the operator
\begin{equation}
\hat{\bar{F}}=U\hat{F}U^{-1}\label{Iop}%
\end{equation}
acting in $L^{2}(Q,\omega_{\bar{g}})$.

\begin{theorem}
\label{main1}Suppose that the operator $\hat{H}$ in the Hilbert space
$L^{2}(Q,\omega_{g})$ is given by (\ref{op1}). Then the operator $U\hat
{H}U^{-1}$, acting in the Hilbert space $L^{2}(Q,\omega_{\bar{g}})$, has the
form%
\begin{equation}
U\hat{H}U^{-1}=-\frac{\hbar^{2}}{2}\bar{\nabla}_{i}A^{ij}\bar{\nabla}%
_{j}+V(x)+\hbar^{2}W(x)\label{W0}%
\end{equation}
with $W(x)$ given by%
\begin{equation}
W(x)=\frac{1}{8}\left[  A^{ij}\left(  \Gamma_{ik}^{k}\Gamma_{js}^{s}%
-\bar{\Gamma}_{ik}^{k}\bar{\Gamma}_{js}^{s}\right)  +2\left(  A^{ij}\left(
\Gamma_{jk}^{k}-\bar{\Gamma}_{jk}^{k}\right)  \right)  _{,i}\right] \label{W2}%
\end{equation}
where the subscript $_{,i}$ denotes differentiation with respect to $x_{i}$.
\end{theorem}

We will call the term $W(x)$ the \emph{quantum correction term} as it
describes what happens to the operator (\ref{op1}) transformed from
$L^{2}(Q,\omega_{g})$ to $L^{2}(Q,\omega_{\bar{g}})$.

\begin{proof}
One can prove this theorem by direct calculations of $U\hat{H}U^{-1}$. Of
course%
\[
U\hat{H}U^{-1}=U\left(  -\frac{\hbar^{2}}{2}\nabla_{i}A^{ij}\nabla
_{j}+V(x)\right)  U^{-1}=-\frac{\hbar^{2}}{2}U\nabla_{i}A^{ij}\nabla_{j}%
U^{-1}+V(x)
\]
By using the fact
\[
\frac{\partial U}{\partial x^{i}}=\frac{1}{2}U\left(  \Gamma_{ik}^{k}%
-\bar{\Gamma}_{ik}^{k}\right)
\]
after some calculations we arrive at (\ref{W0})-(\ref{W2}). Alternatively, the
similarity map (\ref{Iop}) can be calculated using the automorphism $\bar
{S}S^{-1}$. From our general theory \cite{Blaszak:2012}-\cite{Blaszak:2014} it
follows that quantizing the observable $H$ with respect to $g$ yields an
operator that is mapped through (\ref{Iop}) on the operator that we obtain by
quantizing the observable $H^{\prime}=\bar{S}S^{-1}H$ with respect to $\bar
{g}$. This yields, that the operator (\ref{op1}) attains in the space in
$L^{2}(Q,\omega_{\bar{g}})$ the form
\begin{equation}
U\hat{H}U^{-1}=(\bar{S}^{-1}H^{\prime})_{W}(\hat{\bar{x}},\hat{\bar{p}}%
)=(\bar{S}^{-1}\bar{S}S^{-1}H)_{W}(\hat{\bar{x}},\hat{\bar{p}})=(S^{-1}%
H)_{W}(\hat{\bar{x}},\hat{\bar{p}})\label{eq:7}%
\end{equation}
Let us thus explicitly calculate the operator on the right hand side of
(\ref{eq:7}). Due to (\ref{Sg}) and using the fact that $H$ is second order in
momenta (so that the only terms in $S^{-1}$ that act on $H$ are or order up to
$\hbar^{2}$, see Remark \ref{uwaga}), after some calculations we obtain%
\begin{equation}
S^{-1}H=H+\frac{1}{2}\hbar^{2}\left(  \frac{1}{4}A_{\phantom{ij},ij}%
^{ij}+\frac{1}{2}A_{\phantom{ij},i}^{ij}\Gamma_{jk}^{k}+\frac{1}{2}%
A^{ij}\Gamma_{ik,j}^{k}+\frac{1}{4}A^{ij}\Gamma_{ik}^{k}\Gamma_{jl}%
^{l}\right)  =\bar{S}^{-1}H+\hbar^{2}W(x)\nonumber
\end{equation}
with%
\begin{equation}
W(x)=\frac{1}{2}\left[  \frac{1}{2}A_{\phantom{ij},i}^{ij}\left(  \Gamma
_{jk}^{k}-\bar{\Gamma}_{jk}^{k}\right)  +\frac{1}{2}A^{ij}\left(
\Gamma_{ik,j}^{k}-\bar{\Gamma}_{ik,j}^{k}\right)  +\frac{1}{4}A^{ij}\left(
\Gamma_{ik}^{k}\Gamma_{jl}^{l}-\bar{\Gamma}_{ik}^{k}\bar{\Gamma}_{jl}%
^{l}\right)  \right] \label{ziuta}%
\end{equation}
coinciding with $W(x)$ in (\ref{W2}).
\end{proof}

In Appendix we show that (\ref{W2}) can be written in a covariant form as%

\begin{equation}
W(x)=\frac{1}{8}\left(  A_{\phantom{ij};i}^{ij}G^{ks}g_{ks;j}+A^{ij}%
G^{ks}g_{ks;ij}+A^{ij}G_{\phantom{ij};i}^{ks}g_{ks;j}+\frac{1}{4}A^{ij}%
G^{kr}g_{kr;i}G^{sl}g_{sl;j}\right) \label{cov}%
\end{equation}
where the covariant derivatives are taken with respect to the connection
$\bar{\nabla}_{i}$. In what follows we will also need a specification of this
correction term to the following situation: suppose that $G=\frac{1}%
{u}G_{B,\theta}$ (where $u=u(x)$) where the metric $G_{B,\theta}$ is flat and
suppose that $\bar{G}=G_{B,\theta}$. Then the correction term (\ref{cov})
attains the form%
\begin{equation}
W(x)=\frac{n}{8}\left(  A^{ij}\frac{u_{,j}}{u}\right)  _{,i}+\frac{n^{2}}%
{32}\frac{1}{u^{2}}A^{ij}u_{,i}u_{,j}\label{covspec}%
\end{equation}

\section{Separable minimal quantizations of St\"{a}ckel systems\label{SQS}}

Suppose we have a St\"{a}ckel system written in arbitrary Darboux coordinates
$(x,p)$:%
\begin{equation}
H_{r}=\frac{1}{2}p^{T}A_{r}p+V_{r}(x),\quad r=1,\dotsc,n\label{1}%
\end{equation}
Given a metric $g$ we can now perform the minimal quantization of our
St\"{a}ckel system (\ref{1}) as described in the previous section. As a result
we obtain $n$ quantum Hamiltonians%

\begin{equation}
\hat{H}_{r}=-\frac{1}{2}\hbar^{2}\nabla_{i}\left(  T_{r}G\right)  ^{ij}%
\nabla_{j}+V_{r}(x),\quad r=1,\ldots,n\label{2}%
\end{equation}
acting in the Hilbert space $L^{2}(Q,\omega_{g})$, $\omega_{g}=\left\vert \det
g\right\vert ^{1/2}dx$, where $A_{r}=T_{r}G$. Let us rewrite the operators
(\ref{2}) in some separation coordinates $(\lambda,\mu)$ for the classical
St\"{a}ckel system (\ref{1}). We will always assume the conditions $1$-$3$
from Section 2. This also means that $g$ and thus $G$ are diagonal in
separation coordinates. Thus, since $A_{r}$ are diagonal in separation
coordinates, so are $T_{r}$. Calculating covariant derivatives we obtain
\begin{align}
\hat{H}_{r} &  =-\frac{1}{2}\hbar^{2}G^{ii}\left(  T_{r}^{(i)}\partial_{i}%
^{2}+(\partial_{i}T_{r}^{(i)})\partial_{i}-T_{r}^{(i)}\Gamma_{i}\partial
_{i}\right)  +V_{r}(\lambda)\nonumber\\
&  =-\frac{1}{2}\hbar^{2}A_{r}^{ii}\left(  \partial_{i}^{2}+\left(
\frac{\partial_{i}T_{r}^{(i)}}{T_{r}^{(i)}}-\Gamma_{i}\right)  \partial
_{i}\right)  +V_{r}(\lambda)\label{3}%
\end{align}
where $T_{r}^{(i)}\equiv(T_{r})_{i}^{i}$ (no summation) and where $\Gamma_{i}$
are metrically contracted Christoffel symbols (\ref{mcCh}). In orthogonal
coordinates they read \cite{Ben2}
\[
\Gamma_{i}=\frac{1}{2}\frac{\partial_{i}\det G}{\det G}-\frac{\partial
_{i}G^{ii}}{G^{ii}}%
\]

The next theorem, proved in \cite{bletal2}, follows directly from (\ref{3}).

\begin{theorem}
\label{main2}The necessary and sufficient condition for quantum separability
of operators $\hat{H}_{r}$ takes the form
\begin{equation}
\Xi_{i}=\Xi_{i}(\lambda_{i})\text{ or }\partial_{j}\Xi_{i}=0,\quad j\neq
i\label{4}%
\end{equation}
where
\[
\Xi_{i}=\frac{\partial_{i}T_{r}^{(i)}}{T_{r}^{(i)}}-\Gamma_{i}%
\]

\end{theorem}

We will call the condition (\ref{4}) the \emph{generalized Robertson
condition}. Indeed, due to (\ref{Ar}), the operators (\ref{3}) can then be
written as
\begin{equation}
\hat{H}_{r}=-\frac{1}{2}\hbar^{2}\left(  S_{\gamma}^{-1}\right)  _{r}^{i}%
f_{i}(\lambda_{i})\left(  \partial_{i}^{2}+\Xi_{i}(\lambda_{i})\partial
_{i}\right)  +\left(  S_{\gamma}^{-1}\right)  _{r}^{i}\sigma_{i}(\lambda
_{i}),\quad r=1,\dotsc,n\label{4a}%
\end{equation}
and then application of the St\"{a}ckel matrix $S_{\gamma}$ to the system of
eigenvalue problems for (\ref{4a})
\begin{equation}
S_{\gamma}%
\begin{pmatrix}
\hat{H}_{1}\Psi\\
\vdots\\
\hat{H}_{n}\Psi
\end{pmatrix}
=S_{\gamma}%
\begin{pmatrix}
E_{1}\Psi\\
\vdots\\
E_{n}\Psi
\end{pmatrix}
\label{4b}%
\end{equation}
separates (\ref{4b}) to $n$ one-dimensional eigenvalue problems
\begin{equation}
(E_{1}\lambda_{i}^{\gamma_{1}}+E_{2}\lambda_{i}^{\gamma_{2}}+\dotsb+E_{n}%
)\psi_{i}(\lambda_{i})=-\frac{1}{2}\hbar^{2}f_{i}(\lambda_{i})\left[
\frac{d^{2}\psi_{i}(\lambda_{i})}{d\lambda_{i}^{2}}+\Xi_{i}(\lambda_{i}%
)\frac{d\psi_{i}(\lambda_{i})}{d\lambda_{i}}\right]  +\sigma_{i}(\lambda
_{i})\psi_{i}(\lambda_{i}),\quad i=1,\ldots,n\label{4c}%
\end{equation}
called separation equations or quantum separable relations, so that
\[
\Psi(\lambda_{1},\dotsc,\lambda_{n},c,E)=\prod_{i=1}^{n}\psi_{i}(\lambda
_{i},c_{2i-1,}c_{2i},E)
\]
is a common, multiplicatively separable solution of stationary Schr\"{o}dinger
equations for all $\hat{H}_{r}$, satisfying the definition of separability
from \cite{Ben1}. The constants $E_{i}$ are unspecified unless some boundary
conditions are imposed while $c_{2i-1,}c_{2i}$ are integration constants
originating during the process of solving equation $i$ in (\ref{4c}); there
are $2n$ of them in total. In the case $G=A_{1}$ (or, in general, $G$ equal to
any $A_{s}$) $T_{r}$ are Killing tensors of $g$ so in $\lambda$-coordinates
$\partial_{i}T_{r}^{(i)}=0.$ In consequence the condition (\ref{4}) reduces to
the Robertson condition for quantum separability (\ref{rob1}) or (\ref{rob}).

In \cite{bletal} we proved that for the case $G=A_{1}$ the only class of
St\"{a}ckel systems (\ref{St}) for which the Robertson condition (\ref{rob})
is satisfied is the Benenti class where
\begin{equation}
\Gamma_{i}=-\frac{1}{2}\frac{f_{i}^{\prime}(\lambda_{i})}{f_{i}(\lambda_{i}%
)}\label{GammaB}%
\end{equation}
For all other choices of $\gamma_{i}$ in (\ref{St}) this condition fails. In
\cite{bletal2} we investigated the more general case when $G$ is not one of
the tensors $A_{r}$ in (\ref{Ham}) but is a flat metric from the Benenti class
(\ref{GB}). We showed that also in this case the only class of St\"{a}ckel
systems (\ref{St}) that is quantum separable is again the Benenti class. It
means that in order to achieve quantum separability of an arbitrary
St\"{a}ckel system of the type (\ref{St}) we have to consider a broader class
of admissible metric tensors $g$ used in the quantization procedure.

Consider thus a St\"{a}ckel system (\ref{St}) defined by some fixed choice of
$\gamma_{1}>\gamma_{2}>\dotsb>\gamma_{n}=0$ and the choice of $f_{i}%
,\sigma_{i}$. We will now search for the metric $G$ that satisfies the
generalized Robertson condition (\ref{4}) for this St\"{a}ckel system. Due to
the structure (\ref{Are}) of $A_{r}$ we look for $G$ in the form
\begin{equation}
G=u^{-1}(\lambda)G_{B,\theta}\label{met}%
\end{equation}
where $G_{B,\theta}$ is the Benenti metric given by (\ref{GB}) with $n$
arbitrary functions $\theta_{i}(\lambda_{i})$ and where $u$ is some function
on $Q$. Albeit this choice is by no means the most general one it will prove
to be sufficiently general. The tensors $T_{r}\,$\ become in this case%
\[
T_{r}=\frac{u}{\varphi}\chi_{r}G_{B,f}g_{B,\theta}%
\]
where $\varphi$ is again given by (\ref{phi}) and where as usual $g_{B,\theta
}=G_{B,\theta}^{-1}$. Plugging this into (\ref{4}) we get%
\begin{equation}
\frac{\partial_{i}T_{r}^{(i)}}{T_{r}^{(i)}}-\Gamma_{i}=\frac{\kappa
_{i}^{\prime}(\lambda_{i})}{\kappa_{i}(\lambda_{i})},\quad=1,\ldots
,n\label{wk}%
\end{equation}
where $\kappa_{i}$ are arbitrary functions of one variable (the right hand
side is just a convenient for us way of writing an arbitrary function of
$\lambda_{i}$). Since for (\ref{met})%
\[
\Gamma_{i}=\left(  \Gamma_{B,\theta}\right)  _{i}+\left(  1-\frac{1}%
{2}n\right)  \frac{\partial_{i}u}{u}%
\]
with $\left(  \Gamma_{B,\theta}\right)  _{i}$ being the metrically contracted
Christoffel symbols for the metric $G_{B,\theta}$, the formula (\ref{wk})
takes the form%
\[
\frac{n}{2}\frac{\partial_{i}u}{u}-\frac{\partial_{i}\varphi}{\varphi}%
=\frac{\kappa_{i}^{\prime}(\lambda_{i})}{\kappa_{i}(\lambda_{i})}+\frac{1}%
{2}\frac{\theta_{i}^{\prime}(\lambda_{i})}{\theta_{i}(\lambda_{i})}%
-\frac{f_{i}^{\prime}(\lambda_{i})}{f_{i}(\lambda_{i})},\quad=1,\ldots,n
\]
which has a solution%
\begin{equation}
u=\varphi^{\frac{2}{n}}\prod\limits_{i=1}^{n}\left(  \frac{\left\vert
\theta_{i}\right\vert \kappa_{i}^{2}}{f_{i}^{2}}\right)  ^{\frac{1}{n}%
}\label{sol}%
\end{equation}
In order to receive a solution as simple as possible we choose $\kappa_{i}$ so
that
\[
\frac{\left\vert \theta_{i}\right\vert \kappa_{i}^{2}}{f_{i}^{2}}=1
\]
(notice that $\theta_{i}$ are still arbitrary) yielding (\ref{sol}) in the
form $u=\varphi^{\frac{2}{n}}$. Thus, we have proved

\begin{theorem}
\label{wazny}Suppose $\theta_{i},i=1,\ldots,n$ are $n$ arbitrary functions of
one variable. Then applying the procedure of minimal quantization, with the
metric tensor
\begin{equation}
g=\varphi^{\frac{2}{n}}g_{B,\theta}\label{cf}%
\end{equation}
where $g_{B,\theta}=G_{B,\theta}^{-1}$ with $G_{B,\theta}$ given by%
\begin{equation}
G_{B,\theta}=\diag\left(  \frac{\theta_{1}(\lambda_{1})}{\Delta_{1}}%
,\ldots,\frac{\theta_{n}(\lambda_{n})}{\Delta_{n}}\right) \label{theta}%
\end{equation}
to the St\"{a}ckel system (\ref{St}) we obtain a quantum separable system
(\ref{2}) with the separation equations of the form%
\begin{equation}
(E_{1}\lambda_{i}^{\gamma_{1}}+E_{2}\lambda_{i}^{\gamma_{2}}+\dotsb+E_{n}%
)\psi_{i}(\lambda_{i})=-\frac{1}{2}\hbar^{2}f_{i}(\lambda_{i})\left[
\frac{d^{2}\psi_{i}(\lambda_{i})}{d\lambda_{i}^{2}}+\left(  \frac
{f_{i}^{\prime}(\lambda_{i})}{f_{i}(\lambda_{i})}-\frac{1}{2}\frac{\theta
_{i}^{\prime}(\lambda_{i})}{\theta_{i}(\lambda_{i})}\right)  \frac{d\psi
_{i}(\lambda_{i})}{d\lambda_{i}}\right]  +\sigma_{i}(\lambda_{i})\psi
_{i}(\lambda_{i}),\label{sep}%
\end{equation}
where $i=1,\ldots,n.$
\end{theorem}

The metric $g$ in (\ref{cf}) is a \emph{conformal deformation} of the Benenti
metric $g_{B,\theta}$. Thus, there exists an infinite family of separable
quantizations of a St\"{a}ckel system (\ref{St}) parametrized by $n $
arbitrary functions $\theta_{i}$ of one variable: any St\"{a}ckel system
(\ref{St}) can be separably quantized in the conformally deformed metric
(\ref{cf}) (note that this metric is conformally flat in the case when
$g_{B,\theta}$ is flat). Moreover, since for the Benenti class $\varphi=1,$
any St\"{a}ckel system from the Benenti class (\ref{Stb}) can be separably
quantized in any metric of Benenti class (\ref{theta}), including the subclass
of flat metrics.

\section{Quantum integrability of St\"{a}ckel systems in arbitrary Hilbert
spaces\label{QIS}}

We remind the reader that in \cite{Ben2} the authors derived the necessary and
sufficient condition for commutativity of quantum Hamiltonians $\hat{H}_{r}$
of the form (\ref{16}) (and with $A_{1}=G$) called the pre-Robertson condition
(\ref{prerob1}) or (\ref{prerob}), which took the form%
\begin{equation}
\partial_{i}^{2}\Gamma_{j}-\Gamma_{i}\partial_{i}\Gamma_{j}=0,\quad i\neq
j.\label{5}%
\end{equation}
In our case, when $G$ is not related with any $A_{r}$, analogous calculations
lead to the following necessary and sufficient condition for commutativity of
$\hat{H}_{r}$ which we call the generalized pre-Robertson condition
\cite{bletal2}:%
\begin{equation}
\partial_{i}^{2}\Xi_{j}-\Xi_{i}\partial_{i}\Xi_{j}=0,\quad i\neq j.\label{6}%
\end{equation}

Assume that we have a St\"{a}ckel system $H_{r}$, $r=1,\ldots,n$ of the form
(\ref{St}). Let us perform the procedure of minimal quantization of this
system in the metric $G$ given by (\ref{cf}), as described in the previous
section. We obtain then the quantum separable system consisting of $n$
operators $\hat{H}_{r}$ acting on the Hilbert space $L^{2}(Q,\omega_{g})$,
$\omega_{g}=\left\vert \det g\right\vert ^{1/2}d\lambda$. Since the
generalized Robertson condition (\ref{4}) implies the generalized
pre-Robertson condition (\ref{6}) we conclude that this system is also quantum
integrable: $\left[  \hat{H}_{r},\hat{H}_{s}\right]  =0$. Using Theorem
\ref{main1} we are able to write operators $\hat{H}_{r}$ in another metric
$\bar{g}$ i.e. in the Hilbert space $L^{2}(Q,\omega_{\bar{g}})$ which yields
new quantum operators $\hat{\bar{H}}_{r}$, $r=1,\ldots,n$ that constitute
again quantum integrable (but not necessarily quantum separable) system. Due
to the theory developed in Section \ref{MinQ} we know, that we can equally
well take the classical Hamiltonians $H_{r}$ amended by quantum correction
terms, i.e. the functions $H_{r}+\hbar^{2}W_{r}$ with $W_{r}$ given by
(\ref{W2}) (or equivalently by (\ref{cov})) and minimally quantize them in the
metric $\bar{g}$ as this will yield the same quantum integrable system
$\hat{\bar{H}}_{r}$, $r=1,\ldots,n$.

In \cite{stacktr} we demonstrated that any St\"{a}ckel system of the class
(\ref{St}) can be constructed by an appropriate St\"{a}ckel transform of a
suitably chosen flat St\"{a}ckel system from Benenti class. Moreover, in
\cite{flatstackel} we explicitly constructed flat coordinates for any flat
St\"{a}ckel system. Therefore we are able to write down our original
St\"{a}ckel system $H_{r}$, $r=1,\ldots,n$ in flat coordinates of the metric
$\bar{g}$ of the form (\ref{GB}) ( $\bar{g}$ is flat as soon the conditions in
Remark \ref{uwaga2} are satisfied). In this specific case, if we apply the
standard Weyl quantization to the St\"{a}ckel system $H_{r}+\hbar^{2}W_{r}$
(i.e. our original system amended by the quantum correction terms $\hbar
^{2}W_{r}$) we will obtain a quantum integrable system. One can also say,
alternatively, that if we want to avoid quantum correction terms, we should
quantize the original system $H_{r}$, $r=1,\ldots,n$ not by Weyl quantization
but by minimal quantization in a suitably chosen conformally flat metric $G$.

In papers \cite{Hiet1} and \cite{Hiet2} the authors presented some ad hoc
calculations generating quantum correction terms that guarantee integrability
of quantum systems obtained through Weyl quantization of some Hamiltonian
systems. Our theory shows how to construct these quantum correction terms in a
systematic way (albeit within the class of St\"{a}ckel systems, not considered
in \cite{Hiet1}-\cite{Hiet2}). We will illustrate this on two examples below.
It is important to stress that the presented systems cannot be separably
quantized in the frame of the classical Robertson-Eisenhart formalism.

\begin{example}
Consider the St\"{a}ckel system (\ref{St}) \ for $n=3$ given by the separation
relations of the form:%
\begin{equation}
H_{1}\lambda_{i}^{3}+H_{2}\lambda_{i}+H_{3}=\frac{1}{2}\lambda_{i}\mu_{i}%
^{2}+\lambda_{i}^{4},\quad i=1,2,3\label{Hp}%
\end{equation}
so that $\gamma_{1}=3$, $\gamma_{2}=1$ and $\gamma_{3}=0$ and with
$f_{i}(\lambda_{i})=\lambda_{i}$ and $\sigma_{i}(\lambda_{i})=\lambda_{i}^{4}%
$. In this case $\varphi=\rho_{1}(\lambda)=-(\lambda_{1}+\lambda_{2}%
+\lambda_{3})$. Consider also the corresponding metric $G_{B,f}$ given by
(\ref{GB}). This metric is flat, by Remark \ref{uwaga2}. In the coordinates
$x_{1},x_{2},x_{3}$ defined through (cf. \ref{defq}))
\begin{gather}
\rho_{1}\equiv-\left(  \lambda_{1}+\lambda_{2}+\lambda_{3}\right)
=x_{1}\nonumber\\
\rho_{2}\equiv\lambda_{1}\lambda_{2}+\lambda_{1}\lambda_{3}+\lambda_{2}%
\lambda_{3}=x_{2}+\frac{1}{4}x_{1}^{2}\label{plaskie}\\
\rho_{3}\equiv-\lambda_{1}\lambda_{2}\lambda_{3}=-\frac{1}{4}x_{3}%
^{2}\nonumber
\end{gather}
the metric $G_{B,f}$ reads%
\begin{equation}
G_{B,f}=\left(
\begin{array}
[c]{ccc}%
0 & 1 & 0\\
1 & 0 & 0\\
0 & 0 & 1
\end{array}
\right) \label{Gp}%
\end{equation}
so $\varphi=x_{1}$ in $x_{i}$-coordinates and $x_{i}$ are flat non-orthogonal
coordinates for $G_{B,f}$. Solving the relations (\ref{Hp}) with respect to
the Hamiltonians $H_{i}$ and passing to the variables $x_{i}$ we receive
$H_{r}=A_{r}^{ij}y_{i}y_{j}+V_{r}(x)$ where $y_{i}$ are momenta conjugate to
$x_{i}$ and where the tensors $A_{r}$ have the form
\begin{gather*}
A_{1}=%
\begin{pmatrix}
0 & -\frac{1}{x_{1}} & 0\\
-\frac{1}{x_{1}} & 0 & 0\\
0 & 0 & -\frac{1}{x_{1}}%
\end{pmatrix}
,\quad A_{2}=%
\begin{pmatrix}
1 & \frac{1}{4}x_{1}-\frac{x_{2}}{x_{1}} & 0\\
\frac{1}{4}x_{1}-\frac{x_{2}}{x_{1}} & -x_{2} & -\frac{1}{2}x_{3}\\
0 & -\frac{1}{2}x_{3} & \frac{3}{4}x_{1}-\frac{x_{2}}{x_{1}}%
\end{pmatrix}
,\\
A_{3}=%
\begin{pmatrix}
0 & \frac{1}{4}\frac{x_{3}^{2}}{x_{1}} & -\frac{1}{2}x_{3}\\
\frac{1}{4}\frac{x_{3}^{2}}{x_{1}} & \frac{1}{4}x_{3}^{2} & -\frac{1}{4}%
x_{1}x_{3}\\
-\frac{1}{2}x_{3} & -\frac{1}{4}x_{1}x_{3} & \frac{1}{4}x_{1}^{2}+x_{2}%
+\frac{1}{4}\frac{x_{3}^{2}}{x_{1}}%
\end{pmatrix}
\end{gather*}
with the corresponding rational potentials%
\begin{align*}
V_{1}(x) &  =-\frac{3}{4}x_{1}+\frac{x_{2}}{x_{1}}\\
V_{2}(x) &  =\frac{1}{16}x_{1}^{3}+\frac{1}{2}x_{1}x_{2}+\frac{1}{4}x_{3}%
^{2}+\frac{x_{2}^{2}}{x_{1}}\\
V_{3}(x) &  =-\frac{1}{16}x_{1}x_{3}^{2}-\frac{1}{4}\frac{x_{2}x_{3}^{2}%
}{x_{1}}%
\end{align*}
From our theory it follows that we can perform a separable quantization of
this system in the conformally flat metric $G=\frac{1}{u}G_{B,f}$ (which means
that we choose $\theta_{i}=f_{i}$) with $u=\varphi^{2/n}=x_{1}^{2/3}$. We
obtain three commuting operators%
\begin{equation}
\hat{H}_{r}=-\frac{1}{2}\hbar^{2}\nabla_{i}A_{r}^{ij}\nabla_{j}+V_{r}%
(x)\label{opn}%
\end{equation}
(where $\nabla_{i}$ is the connection defined by $G$),acting in the Hilbert
space $L^{2}(Q,\omega_{g})=L^{2}(Q,\left\vert x_{1}\right\vert dx)$
($\omega_{g}=\left\vert \det g\right\vert ^{1/2}dx=\left\vert u^{3/2}%
\right\vert dx=\left\vert x_{1}\right\vert dx$). In the separation coordinates
$(\lambda,\mu)$ the separation equations (\ref{sep}) for $\hat{H}_{r}$ attain
the form%
\begin{equation}
(E_{1}\lambda_{i}^{3}+E_{2}\lambda_{i}+E_{3})\psi_{i}(\lambda_{i})=-\frac
{1}{2}\hbar^{2}\left(  \lambda_{i}\frac{d^{2}\psi_{i}(\lambda_{i})}%
{d\lambda_{i}^{2}}+\frac{1}{2}\frac{d\psi_{i}(\lambda_{i})}{d\lambda_{i}%
}\right)  +\lambda_{i}^{4}\psi_{i}(\lambda_{i}),\quad i=1,2,3\label{S}%
\end{equation}
Let us now rewrite our operators (\ref{opn}) in the Hilbert space
$L^{2}(Q,\omega_{\bar{g}})=L^{2}(Q,dx)$ ($\omega_{\bar{g}}=\left\vert \det
\bar{g}\right\vert ^{1/2}dx=dx$) with the flat metric $\bar{G}=G_{B,f}$. From
our theory it follows that a suitable way to do it is to quantize our
Hamiltonians $H_{r}$ directly in the metric $\bar{G}$ after amending them by
the quantum correction terms $W_{i}(x)$ given by (\ref{covspec})%
\[
W_{1}=0\text{, }W_{2}=-\frac{3}{8}\frac{1}{x_{1}^{2}}\text{, }W_{3}=-\frac
{1}{8}\frac{1}{x_{1}}%
\]
One can check by direct calculations that the operators
\begin{equation}
\hat{\bar{H}}_{r}=-\frac{1}{2}\hbar^{2}\partial_{i}A_{r}^{ij}\partial
_{j}+\hbar^{2}W_{r}(x)+V_{r}(x),\quad r=1,\ldots,n\label{opp}%
\end{equation}
%}
(the coordinates $x_{i}$ are flat for $\bar{g}=G_{B,f}$ so $\bar{\nabla}%
_{i}=\partial_{i}=\partial/\partial x_{i}$) do indeed commute, thus
constituting a quantum integrable system. The operators (\ref{opp}) are
however not quantum separable, contrary to the operators (\ref{opn}), but are
R-separable. It means that in separation coordinates
\[
\widehat{\bar{H}}_{r}\bar{\Psi}(\lambda)=E_{r}\bar{\Psi}(\lambda),\quad
\bar{\Psi}(\lambda)=U(\lambda)\Psi(\lambda)=(\lambda^{1}+\lambda^{2}%
+\lambda^{3})^{\frac{1}{2}}\psi(\lambda^{1})\psi(\lambda^{2})\psi(\lambda
^{3}),
\]
and $\psi(\lambda^{i})$ solves (\ref{S}).
\end{example}

\begin{example}
In our second example we consider the following St\"{a}ckel system%
\begin{equation}
H_{1}\lambda_{i}^{3}+H_{2}\lambda_{i}^{2}+H_{3}=\frac{1}{2}\lambda_{i}\mu
_{i}^{2}+\lambda_{i}^{4},\quad i=1,2,3\label{Hp2}%
\end{equation}
so that this time $\gamma_{1}=3$, $\gamma_{2}=2$ and $\gamma_{3}=0$ but still
with $f_{i}(\lambda_{i})=\lambda_{i}$ and $\sigma_{i}(\lambda_{i})=\lambda
_{i}^{4}$. In this case $\varphi=\rho_{2}(\lambda)=\lambda_{1}\lambda
_{2}+\lambda_{1}\lambda_{3}+\lambda_{2}\lambda_{3}$. We consider again the
same metric $G_{B,f}$ with the same flat coordinates $x_{i}$ given by
(\ref{plaskie}). This time the tensors $A_{r}$ have the form%
\begin{gather*}
A_{1}=\frac{1}{\rho_{2}(x)}\left(
\begin{array}
[c]{ccc}%
-1 & -\frac{1}{2}\,x_{{1}} & 0\\
\noalign{\medskip}-\frac{1}{2}\,x_{{1}} & x_{{2}} & \frac{1}{2}\,x_{{3}}\\
\noalign{\medskip}0 & \frac{1}{2}\,x_{{3}} & -x_{{1}}%
\end{array}
\right)  ,\quad A_{2}=\frac{1}{\rho_{2}(x)}\left(
\begin{array}
[c]{ccc}%
-x_{{1}} & -\frac{1}{4}\,{x_{{1}}}^{2}+x_{{2}} & 0\\
\noalign{\medskip}-\frac{1}{4}\,{x_{{1}}}^{2}+x_{{2}} & x_{{2}}x_{{1}} &
\frac{1}{2}\,x_{{1}}x_{{3}}\\
\noalign{\medskip}0 & \frac{1}{2}\,x_{{1}}x_{{3}} & -\frac{3}{4}\,{x_{{1}}%
}^{2}+x_{{2}}%
\end{array}
\right) \\
A_{3}=\frac{1}{4\rho_{2}(x)}\left(
\begin{array}
[c]{ccc}%
{x_{{3}}}^{2} & \frac{1}{2}\,{x_{{3}}}^{2}x_{{1}} & -\frac{1}{2}\,\left(
{x_{{1}}}^{2}+4\,x_{{2}}\right)  x_{{3}}\\
\noalign{\medskip}\frac{1}{2}\,{x_{{3}}}^{2}x_{{1}} & \frac{1}{4}\,{x_{{3}}%
}^{2}{x_{{\ 1}}}^{2} & -\frac{1}{4}\,x_{{3}}\left(  {x_{{1}}}^{3}+4\,x_{{2}%
}x_{{1}}+2\,{x_{{\ 3}}}^{2}\right) \\
\noalign{\medskip}-\frac{1}{2}\,\left(  {x_{{1}}}^{2}+4\,x_{{2}}\right)
x_{{3}} & -\frac{1}{4}\,x_{{3}}\left(  {x_{{1}}}^{3}+4\,x_{{2}}x_{{1}%
}+2\,{x_{{3}}}^{2}\right)  & \frac{1}{4}\,{x_{{1}}}^{4}+2\,{x_{{1}}}^{2}%
x_{{2}}+4\,{x_{{2}}}^{2}+{x_{{3}}}^{2}x_{{1}}%
\end{array}
\right)
\end{gather*}
where $\rho_{2}(x)=x_{2}+\frac{1}{4}x_{1}^{2}$, while the potentials are%
\begin{align*}
V_{1}(x) &  =-\frac{1}{4\rho_{2}(x)}\left(  x_{1}^{3}+4x_{1}x_{2}+x_{3}%
^{2}\right) \\
V_{2}(x) &  =-\frac{1}{4\rho_{2}(x)}\left(  \frac{1}{4}x_{1}^{4}+2x_{1}%
^{2}x_{2}+x_{1}x_{3}^{2}+4x_{2}^{2}\right) \\
V_{3}(x) &  =\frac{x_{3}^{4}}{16\rho_{2}(x)}%
\end{align*}
This time we perform a separable quantization in the conformally flat metric
$G=\frac{1}{u}G_{B,f}$ with $u=\varphi^{2/n}=\left(  x_{2}+\frac{1}{4}%
x_{1}^{2}\right)  ^{2/3}$. We obtain again three commuting operators%
\begin{equation}
\hat{H}_{r}=-\frac{1}{2}\hbar^{2}\nabla_{i}A_{r}^{ij}\nabla_{j}+V_{r}%
(x)\label{pozno}%
\end{equation}
(where $\nabla_{i}$ is the connection defined by $G$),acting in the Hilbert
space $L^{2}(Q,\left\vert \rho_{2}(x)\right\vert dx)$, while the separation
equations (\ref{sep}) for $\hat{H}_{r}$ become%
\begin{equation}
(E_{1}\lambda_{i}^{3}+E_{2}\lambda_{i}^{2}+E_{3})\psi_{i}(\lambda_{i}%
)=-\frac{1}{2}\hbar^{2}\left(  \lambda_{i}\frac{d^{2}\psi_{i}(\lambda_{i}%
)}{d\lambda_{i}^{2}}+\frac{1}{2}\frac{d\psi_{i}(\lambda_{i})}{d\lambda_{i}%
}\right)  +\lambda_{i}^{4}\psi_{i}(\lambda_{i}),\quad i=1,2,3\label{R1}%
\end{equation}
with the same right hand side as in the previous example. Rewriting our
operators (\ref{opn}) in the Hilbert space $L^{2}(Q,dx)$ with quantization
defined by the flat metric $\bar{G}=G_{B,f}$ leads to the following correction
terms $W_{i}(x)$%
\begin{align*}
W_{1} &  =\frac{1}{16\rho_{2}^{3}(x)}\left(  5x_{1}^{2}-4x_{2}\right) \\
W_{2} &  =\frac{1}{32\rho_{2}^{3}(x)}\left(  7x_{1}^{3}-20x_{1}x_{2}\right) \\
W_{3} &  =-\frac{1}{128\rho_{2}^{3}(x)}\left(  x_{1}^{5}+8x_{1}^{3}%
x_{2}+13x_{1}^{2}x_{3}^{2}+16x_{1}x_{2}^{2}+4x_{2}x_{3}^{2}\right)
\end{align*}
Again, the operators
\begin{equation}
\hat{\bar{H}}_{r}=-\frac{1}{2}\hbar^{2}\partial_{i}A_{r}^{ij}\partial
_{j}+\hbar^{2}W_{r}(x)+V_{r}(x),\quad r=1,\ldots,n\label{oper}%
\end{equation}
commute, as it can be checked for example in Maple. Operators (\ref{oper}) are
R-separable and in separation coordinates
\[
\widehat{\bar{H}}_{r}\bar{\Psi}(\lambda)=E_{r}\bar{\Psi}(\lambda),\quad
\bar{\Psi}(\lambda)=U(\lambda)\Psi(\lambda)=(\lambda^{1}\lambda^{2}%
+\lambda^{1}\lambda^{3}+\lambda^{2}\lambda^{3})^{\frac{1}{2}}\psi(\lambda
^{1})\psi(\lambda^{2})\psi(\lambda^{3}),
\]
where $\psi(\lambda^{i})$ solves (\ref{R1}).
\end{example}

\section{Appendix}

We sketch here the proof of the fact that formulas (\ref{W2}) and (\ref{cov})
are equivalent. We want to demonstrate that%

\begin{equation}
W(x)=\frac{1}{8}\left[  A^{ij}\left(  \Gamma_{ik}^{k}\Gamma_{js}^{s}%
-\bar{\Gamma}_{ik}^{k}\bar{\Gamma}_{js}^{s}\right)  +2\left(  A^{ij}\left(
\Gamma_{jk}^{k}-\bar{\Gamma}_{jk}^{k}\right)  \right)  _{,i}\right]
\label{Ap11}%
\end{equation}
or, equivalently%

\begin{equation}
W(x)=\frac{1}{2}\left[  \frac{1}{2}A_{\phantom{ij},i}^{ij}\left(  \Gamma
_{jk}^{k}-\bar{\Gamma}_{jk}^{k}\right)  +\frac{1}{2}A^{ij}\left(
\Gamma_{ik,j}^{k}-\bar{\Gamma}_{ik,j}^{k}\right)  +\frac{1}{4}A^{ij}\left(
\Gamma_{ik}^{k}\Gamma_{jl}^{l}-\bar{\Gamma}_{ik}^{k}\bar{\Gamma}_{jl}%
^{l}\right)  \right] \label{Ap12}%
\end{equation}
coincides with%

\begin{equation}
W(x)=\frac{1}{8}\left(  A_{\phantom{ij};i}^{ij}G^{ks}g_{ks;j}+A^{ij}%
G^{ks}g_{ks;ij}+A^{ij}G_{\phantom{ij};i}^{ks}g_{ks;j}+\frac{1}{4}A^{ij}%
G^{kr}g_{kr;i}G^{sl}g_{sl;j}\right) \label{Ap2}%
\end{equation}
where the covariant differentiation is taken with respect to the metric
$\bar{g}$. To this end, denote by $M_{j}^{i}$ the quotient of $\bar{g}_{ij}$
and $g_{ij}$:
\begin{equation}
\bar{g}_{ij}=M_{i}^{k}g_{kj},\label{eq:1}%
\end{equation}
(as such, it is a $(1,1)$-tensor), which yields
\[
G^{ij}=(M^{-1})_{k}^{i}\bar{g}^{kj},
\]
The Christoffel symbols $\Gamma_{jk}^{i}$ can now be expressed through
$\bar{\Gamma}_{jk}^{i}$ in the following way
\begin{align*}
\Gamma_{jk}^{i} &  =\frac{1}{2}G^{il}\left(  g_{lj,k}+g_{lk,j}-g_{jk,l}\right)
\\
&  =\frac{1}{2}(M^{-1})_{r}^{i}\bar{g}^{rl}\left(  M_{l,k}^{s}\bar{g}%
_{sj}+M_{l}^{s}\bar{g}_{sj,k}+M_{l,j}^{s}\bar{g}_{sk}+M_{l}^{s}\bar{g}%
_{sk,j}-M_{j,l}^{s}\bar{g}_{sk}-M_{j}^{s}\bar{g}_{sk,l}\right)
\end{align*}
yielding
\begin{align*}
\Gamma_{jk}^{i} &  =\bar{\Gamma}_{jk}^{i}+\frac{1}{2}(M^{-1})_{r}^{i}%
M_{l,k}^{s}\bar{g}^{rl}\bar{g}_{sj}+\frac{1}{2}(M^{-1})_{r}^{i}M_{l,j}^{s}%
\bar{g}^{rl}\bar{g}_{sk}-\frac{1}{2}(M^{-1})_{r}^{i}M_{j,l}^{s}\bar{g}%
^{rl}\bar{g}_{sk}\\
&  \quad{}-\frac{1}{2}(M^{-1})_{r}^{i}M_{j}^{s}\bar{g}^{rl}\bar{g}%
_{sk,l}+\frac{1}{2}\bar{g}^{is}\bar{g}_{jk,s}.
\end{align*}
Using
\begin{align*}
0 &  =\bar{g}_{jk;s}=\bar{g}_{jk,s}-\bar{g}_{nk}\bar{\Gamma}_{js}^{n}-\bar
{g}_{jn}\bar{\Gamma}_{ks}^{n},\\
M_{l,k}^{s} &  =M_{l;k}^{s}-M_{l}^{n}\bar{\Gamma}_{nk}^{s}+M_{n}^{s}%
\bar{\Gamma}_{lk}^{n}%
\end{align*}
(where $;$ denotes the covariant differentiation with respect to $\bar{g}$) we
receive
\[
\Gamma_{jk}^{i}=\bar{\Gamma}_{jk}^{i}+\frac{1}{2}(M^{-1})_{r}^{i}M_{j;k}%
^{r}+\frac{1}{2}(M^{-1})_{r}^{i}M_{k;j}^{r}-\frac{1}{2}(M^{-1})_{r}^{i}%
M_{j;l}^{s}\bar{g}^{rl}\bar{g}_{sk}.
\]
In particular
\[
\Gamma_{jk}^{k}=\bar{\Gamma}_{jk}^{k}+\frac{1}{2}(M^{-1})_{r}^{k}M_{k;j}^{r}.
\]
Moreover%
\[
A_{\phantom{ij},i}^{ij}=A_{\phantom{ij};i}^{ij}-\bar{\Gamma}_{si}^{j}%
A^{is}-\bar{\Gamma}_{si}^{i}A^{sj}%
\]
Inserting all this into (\ref{Ap12}) we obtain
\[
W=\frac{1}{8}\left(  A_{\phantom{ij};i}^{ij}(M^{-1})_{r}^{k}M_{k;j}^{r}%
+A^{ij}(M^{-1})_{r}^{k}M_{k;ij}^{r}+A^{ij}(M^{-1})_{r;i}^{k}M_{k;j}^{r}%
+\frac{1}{4}A^{ij}(M^{-1})_{r}^{k}M_{k;i}^{r}(M^{-1})_{s}^{l}M_{l;j}%
^{s}\right)
\]
that due to (\ref{eq:1}) coincides with (\ref{Ap2}).

\section*{Acknowledgments}

Z.~Doma\'nski acknowledge the support of Polish National Science Center grant
under the contract number DEC-2011/02/A/ST1/00208.

\end{document}